\documentclass[nologo,url,11pt,a4paper]{ETHpaper}

\usepackage{graphicx}                  

\title{Systemic risk in a network fragility model analyzed with probability density evolution of persistent random walks}

\author{Jan Lorenz, Stefano Battiston}

\address{Chair of Systems Design, ETH  Zurich, Kreuzplatz 5, 8032 Zurich, Switzerland \\
  \url{jalorenz@ethz.ch}, \url{sbattiston@ethz.ch}}

\www{\url{http://www.sg.ethz.ch}}

\usepackage{amsmath}
\usepackage{amsfonts}
\usepackage{array}
\usepackage{amsthm}

\newtheorem{proposition}{Proposition}
\newcommand{\sign}{\mathrm{sign}}
\newcommand{\Prob}{\mathrm{Pr}}
\newcommand{\R}{\mathbb{R}}

\newcommand{\one}{\mathbb{\bf 1}}
\newcommand{\tr}{\mathrm{tr}}

\begin{document}

\maketitle

\begin{abstract}
  We study the mean field approximation of a recent model of cascades on
  networks relevant to the investigation of systemic risk control in
  financial networks. In the model, the hypothesis of a trend
  reinforcement in the stochastic process describing the fragility of the
  nodes, induces a trade-off in the systemic risk with respect to the
  density of the network. Increasing the average link density, the
  network is first less exposed to systemic risk, while above an
  intermediate value the systemic risk increases. This result offers a
  simple explanation for the emergence of instabilities in financial
  systems that get increasingly interwoven.  In this paper, we study the
  dynamics of the probability density function of the average fragility.
  This converges to a unique stable distribution which can be computed
  numerically and can be used to estimate the systemic risk as a function
  of the parameters of the model.

\end{abstract}

\section{Introduction}
\label{sec:motivation}

\subsection{Systemic Risk in Financial Networks}
A network of interdependent units which, individually, are susceptible to
fail, is potentially exposed to multiple joint failures of a significant
fraction of units in the system. This is the notion that is usually
associated with the term systemic risk. Systemic risk is particularly
important in the context of infrastructure networks, such as power grids,
and in financial networks. These latter should be meant in a broad sense,
including units of different types, such as business firms, insurance
companies, banks, mutual funds and other financial institutions that are
linked by credit relationships. For instance, if one or more firms fail
and are not able to pay back their debts to the bank, this affect the
balance sheet of the bank which might try to improve its own situation by
increasing the interest rate to the other firms, causing other failures
among the firms. If finally the bank itself fails, this affects
negatively the banks that are linked to it by interbank loans. This is
somehow similar to failure cascades in power grids where a failing power
line implies a higher load an other lines which might bring them to fail.
The size distribution of such failure avalanches is one way of
quantifying the systemic risk.

There is a growing body of literature in economics on financial networks,
that investigates also the issue of systemic risk. While banks-firms
credit relationships have been extensively studied (for an overview, see
\cite{stiglitz_greenwald2003new_paradigm}), only recent works have
analysed phenomena of financial contagion in interbank credit
\cite{allen_gale01financial,freixas_parigi_rochet_00_systemic_risk} and
trade credit. The latter, is a form of credit among business firms,
typically in a supplier-customer relation, which has been less
investigated despite the fact that in some countries it represents a
significant part of the short-term liabilities of the corporate sector
\cite{boyssay06credit_chains}. In the literature on complex networks only
few works have dealt with financial networks, mainly in the context of
self-organized criticality \cite{aleksiejuk_holyst01_soc_banks,
  iori_jafarey_padilla_06_systemic_risk}. Most of those works suggest
that when the degree of the nodes in the network increases the network is
less exposed to systemic risk. In some cases, the evidence that systemic
failures may more rare but also more severe has been found (see for
instance \cite{iori_jafarey_padilla_06_systemic_risk}).

\subsection{The Fragility Model for Cascades on Networks}
In this paper, we consider the model of cascades on networks introduced
by \cite{battiston_al_2007_cascade_on_networks}, in which a clear
tradeoff emerges in the systemic risk, as a function of the network
density. This means that up to an intermediate level of network density
there is a benefit in creating links between units because they allow to
diversify the risk. However, above a certain level of density, the
probability of many joint failures increases. This effect depends on the
presence of a sort of trend reinforcing term in the dynamics of the
fragility of the nodes. The fragility is a state variable that determines
the failure of the node, when it exceeds a given threshold, as well as
subsequent transfer of some damage to the connected nodes. The trend
reinforcing of the fragility corresponds to the following idea.  If the
fragility of a firm at the end of the year has reduced compared to last
year, the firm is rated better in terms of solvency and it has easier
access to credit.  Conversely, if the fragility has increased, the firm
faces worse conditions for credits and thus additional cost that are
likely to increase its fragility furthermore. Notice that, through the
links in the network, this propagates also to the neighbours, since the
fragility of the firm affects the fragility of the neighbours. For
instance, hedge funds leverage even small differences in performance
across firms by 'short-selling' the stocks of the slightly worse ones and
'going long' on the slightly better ones. Thus, even small differences in
the evolution of two firms may matter a lot. Further on, effects like
predatory trading \cite{Brunnermeier.ea05} may induce trend reinforcing.

\subsection{Outline of the Paper}
In \cite{battiston_al_2007_cascade_on_networks}, some analytical results
supporting the simulations are found, based on separating the process of
the evolution of fragility (approximated as a time-dependent
Ornstein-Uhlenbeck process) and the cascade process (where the size of an
avalanche is expressed as the fix point of an equation for the number of
failures). Here, we provide an alternative analysis of the tradeoff
regarding systemic risk mentioned above.  We consider the stochastic
process defined by the mean field approximation of the fragility of the
individual node. This is now a stochastic process for a single variable,
and it is also clear that, having reduced the system to one single
variable, the cascading part of the process is excluded by construction.
In this approximation the failure probability can be taken as a proxy for
systemic risk. In fact, the mean field approximation is valid when all
units behave in a similar way. We study some mathematical properties of
the process and we provide a simple method to show the existence of a
tradeoff in systemic risk as function of the density of the network. The
method is based on recognizing that the process is a combination of a
Gaussian Random Walk (RW) and a Persistent Random Walk (PRW). PRW
\cite{Weiss02} is a variant of the classic RW in which the walker has a
probability $p$ to keep the direction of his former movement and $1-p$ to
switch direction. The process is sometimes called correlated random walk.
It is approximated by the Telegraph's equation
\cite{taylor1922dcm,goldstein1951ddm} in the limit of continuous time and
space. It differs from RW in the scaling with time of the variance of the
displacement of the walker. In our model, the dynamics in time of the
fragility induces a dynamics on the probability density function of its
values.  This dynamics has an exact analytical expression and the
systemic risk is measured as the number of failures in the stable
distribution of fragility. It is possible to proove the existence,
uniqueness and convergence to a stable distribution, based on the
Birkoff-Jentzsch theorem which extends the Perron-Frobenius Theorem to
infinite dimensional vector spaces. We cannot provide an closed-form
expression of the sytemic risk as a function of the parameters of the
model, but we compute the systemic risk numerically, by iterating the
dynamics on the pdf. We show in this way that the systemic risk has
indeed a minimum as function of the network density.

The paper is organized as follows. In Section \ref{sec:model} we
introduce the model. In Section \ref{sec:model_analysis} we analyze the
model: first, we describe the mean-field approximation of the dynamics
and we show how it can be described by using a PRW. Then in Section
\ref{subsec:pdf_dynamics} we derive the dynamics on the probability
density function and we prove existence and uniqueness of the stable pdf.
In Section \ref{sec:numerical_results} we report the results of the
numerical computation of systemic risk. In Section \ref{sec:robustness}
we check the robustness of our results with respect to the type of noise
that enter in the stochastic process of the fragility and some other
slight modifications. In Section \ref{sec:conclusions} we summarize the
results and we draw some conclusions.

\section{The model}
\label{sec:model}

In this section, we describe the network fragility model introduced in
\cite{battiston_al_2007_cascade_on_networks}. Consider a set of $n$ firms
connected in a network, each associated with two state variables, the
size $a$ and the fragility $\varphi$. The first captures the notion of a
proxy for the size of the firm, such as its output. The fragility
captures the notion of financial fragility of the firm. This is measured
for instance in terms of its net worth: when the net worth decreases down
to zero, the firm is not able to pay back its debts and goes bankrupt. So
the larger the net worth, the smaller the fragility. As shown in
\cite{battiston_al_2007_credit_chains}, in a network of firms linked by
supply-customer relationships, the net worth of a firm evolves as a
stochastic process that depends on the net worth of the neighboring
firms. The interaction with the neighbors results in an averaging term
and in a trend reinforcing term. Each firm has a portfolio of suppliers
and customers, which reduces the impact of the fluctuations of prices and
shocks both from the suppliers and customers, thus resulting in the
averaging term. On the other hand, if the production cost increases when
the net worth of the firm and its neighborhood is decreasing (because it
is more costly for the firm to access the credit it need for production),
this results in a trend reinforcing term
\cite{battiston_al_2007_trade_networks_systemic_risk}. Following
\cite{battiston_al_2007_cascade_on_networks} we model directly the
fragility of firms as a stochastic process confined in the interval
$[0,\theta]$, where $\theta$ is the failure threshold.


Firms are connected in a weighted and directed graph with adjacency
matrix $W \in \R^{n\times n}$. $W$ is non-negative and row-stochastic
(i.e. $\sum_j W_{ij}=1$).

As a first step, let us look at the following equation for the evolution
of the fragility of the set of firms

\begin{equation}
  \varphi(t+1) = W\varphi(t) = W^t\varphi(0)
  \label{eq:dyn_only_average}
\end{equation}
where $\varphi=[\varphi_1,\dots,\varphi_n]$ is the vector of fragility
values. If $W_{ij}$ is positive, then the fragility of firms $j$
contributes to a fraction $W_{ij}$ to the value in the next time step of
the fragility of firm $i$.  In other words, the fragility of firm $i$ at
time $t+1$ is a weighted arithmetic mean of the fragility values of the
neighboring firms.  Under some conditions about connectivity in the
network, the values of fragility of the firms will converge in time to a
same value---namely if the matrix $W$ has only one essential class of
indices which is primitive (this is shown in \cite{Seneta73}, such
matrices are called regular if they are row-stochastic, as in our case).
If there are more then one essential classes the fragilities in these
classes converge internally to the same value, as well as all inessential
firms which have connections exclusively to this essential class. But
there is no interplay with fragilities in other essential or inessential
classes. If an essential class is not primitive there is some internal
cycling of fragility values. See \cite{DeGroot74,Berger81} for the
results in the context of conditions of finding consensus in a group of
experts. So, for graph with high link density we could assume that the
fragility values will converge to the same value.


We now introduce additive stochastic shocks and trend reinforcing.
\begin{equation}
  \varphi(t+1) = W(\varphi(t)+\sigma\xi(t)) + \alpha\sign(W(\varphi(t)-\varphi(t-1)))
  \label{eq:dyn}
\end{equation}
In the equation above $\xi(t)$ is a vector of iid random variables,
$\xi_1(t),\dots,\xi_n(t)$, drawn from a distribution $f_\xi$, with
expected value zero and standard deviation one and no skewness (i.e. its
probability density function is symmetric). The parameter $\sigma$
determines the \emph{standard deviation of shocks} and is also called the
\emph{noise level}. The fragility of each firm receives, as a net shock,
the weighted average of the shocks that hit the fragility of the firms in
its neighborhood. In other words, the firm hedges the risk for upward
shocks to its own fragility, by sharing the shocks with other firms. In
the second term of the equation, the $\sign$ is applied component-wise
(for completeness we define $\phi(-1) = 0$) and $\alpha$ is a constant
that we call the \emph{trend strength}. A fixed constant $\alpha$ is
added if the difference between the current average fragility in the
neighborhood and that at the previous time step is positive (i.e. if
fragility has increased) and is subtracted if the difference is negative
(if fragility has decreased).

As a result of the dynamics of Eq. (\ref{eq:dyn}), the values of
fragility may very well go out of the interval $[0,\theta]$. Therefore,
$\phi_i(t+1)$ is set to zero if $\phi_i(t+1) \notin [0,\theta]$ . For
firms whose fragility would go below zero this means that their fragility
cannot become lower than that. For firms that get above $\theta$ this
means that they go bankrupt and are replaced by a new firm with initial
fragility zero. So, Eq. (\ref{eq:dyn}) can be stated as
\begin{displaymath}
  \varphi(t+1) = \one_{[0,\theta]}\left(W(\varphi(t)+\sigma\xi(t)) +
    \alpha\sign(W(\varphi(t)-\varphi(t-1)))\right)
  \label{eq:dyn_chi}
\end{displaymath}
where $\one_{[0,\theta]}$ is the (componentwise) indicator function (e.g.
$\one_{[0,\theta]}(\varphi) = 1$ if $\varphi\in [0,\theta]$ and 0
otherwise, also known as $\chi_{[0,\theta]}$).

In the following we will omit $\one_{[0,\theta]}$ when we describe
dynamics because the reset to zero when a firm fails is not the only
reasonable choice. We discuss some variations at the end of the paper. In
any case throughout we assume that the the process is somewhere reset
when it gets out of $[0,\theta]$.

In the original model in \cite{battiston_al_2007_cascade_on_networks},
when a firm $i$ goes bankrupt, some damage, proportional to the size
$a_i$ of the firm is transferred to the fragility of neighbors. If, as
result, the fragility of some neighbors exceed the threshold $\theta$,
they, in turn, transfer a damage to their (surviving) neighbors. This
cascading process occurs at a faster time scale than the dynamics above.
In this paper, we do not use at all the cascading part of the model. So
Eq. (\ref{eq:dyn_chi}) describes completely the dynamics we study here.

\section{Model analysis}
\label{sec:model_analysis}

Since the dynamics depends on the relative magnitude of the parameters
$\alpha$, $\sigma$ and $\theta$. we can fix $\theta = 1$ without loss of
generality. For abbreviation we define the difference $\Delta\varphi(t) =
\varphi(t)-\varphi(t-1)$.


If $W$ is the unit matrix (i.e. there is no hedging of risk)
\eqref{eq:dyn} reduces to
\begin{equation}
  \varphi_i(t+1) = \varphi_i(t)+\sigma\xi_i(t) +
  \alpha\sign(\Delta\varphi_i(t))
  \label{eq:dyni}
\end{equation}
for all $i$.

If firms are connected in a complete graph and share their fragility
shock to an equal proportion with all other firms, then
$W_{ij}=\frac{1}{n}$ for all $i,j$. In this case, the fragility of each
firm, evolves as the average
$\phi(t)=\frac{1}{n}\sum_{i=1}^k\varphi_i(t)$.
Then, the central limit theorem implies that
\[
\phi(t+1) = \phi(t)+\frac{\sigma}{\sqrt{n}}\xi(t) +
\alpha\sign(\Delta\phi(t)).
\]

In general, if each firm is connected, on average, to $k\leq n$ other
firms, one can make a mean-field approximation of the dynamics of the
fragility of each firm and write
\begin{equation}
  \phi(t+1) = \phi(t)+\frac{\sigma}{\sqrt{k}}\xi(t) +
  \alpha\sign(\Delta\phi(t)).
  \label{eq:dynmf_k}
\end{equation}
The parameter $k$ is the \emph{average number of hedging partners} or
\emph{hedging level}. In other words, the stochastic process on $\phi$
represents the evolution of the average fragility of the economy where
each firm has on average $k$ hedging partners. In this approximation,
increasing the average number of hedging partners $k$ decreases the
standard deviation of the shocks $\sigma$ by a factor of $\sqrt{k}$.
Intuitively, one can expect that the failures becomes less frequent,
because, the smaller are shocks at each time step, the longer it takes to
eventually hit the threshold $\theta$. However, if the noise level
$\sigma$ is very small compared to the trend strength $\alpha$, the
second term in Eq. (\ref{eq:dynmf}) dominates. In particular, if the
fragility was increasing from time $t-1$ to time $t$, then the second
term is for sure equal to $+\alpha$ while the first is probably very
small and therefore the fragility will also increase at time $t+1$.
Therefore, the noise level or equivalently, the average number of
neighbors in the network, seems to play a crucial role for the
probability of a given firm to hit the fragility threshold.

\begin{figure}
  \begin{center}
    \includegraphics[width=0.49\textwidth]{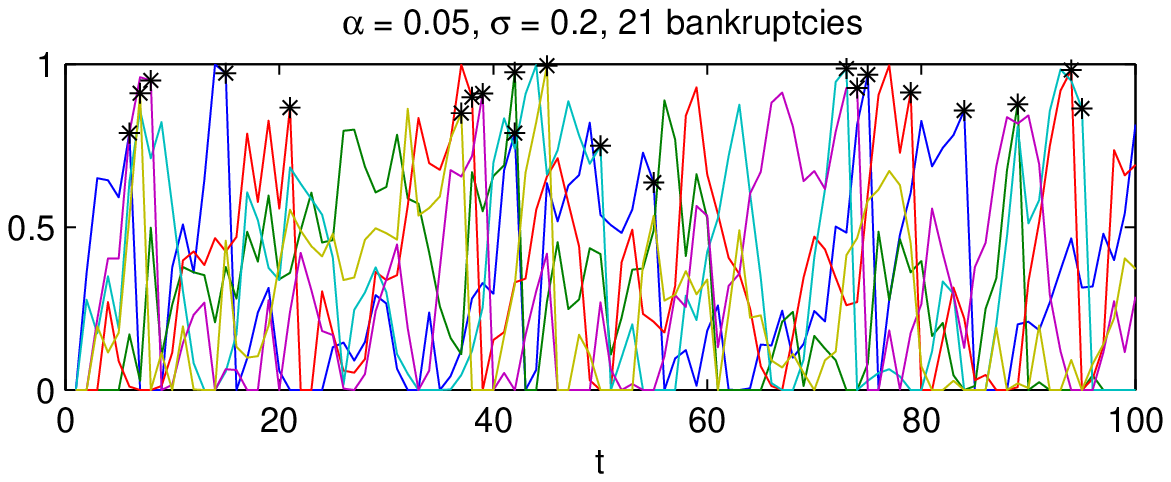}
    \includegraphics[width=0.49\textwidth]{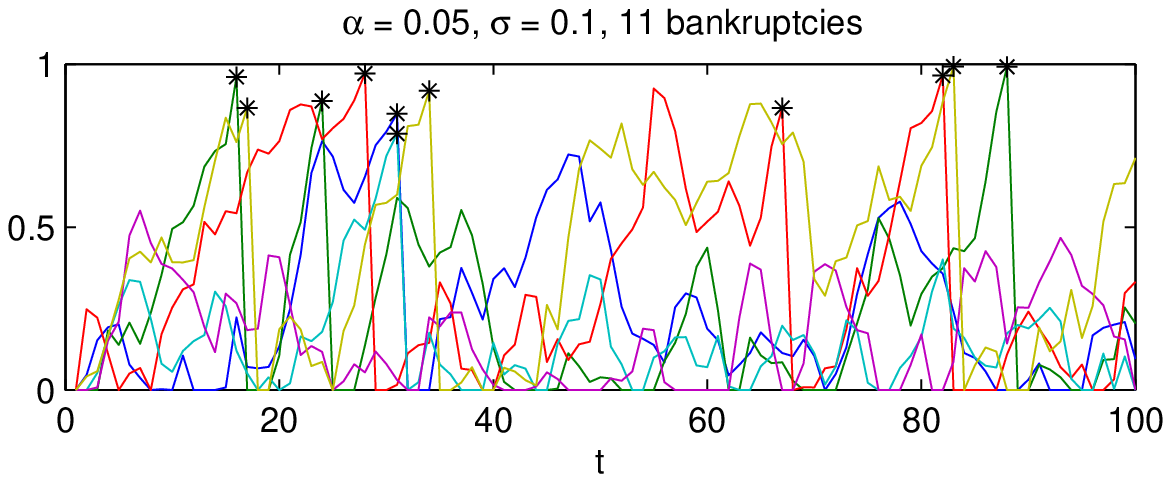}
    \includegraphics[width=0.49\textwidth]{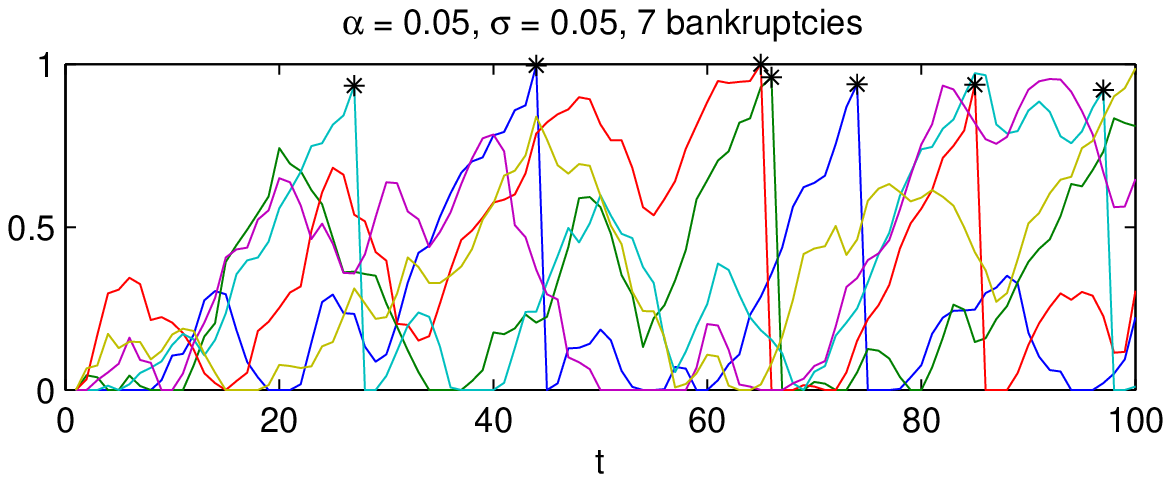}
    \includegraphics[width=0.49\textwidth]{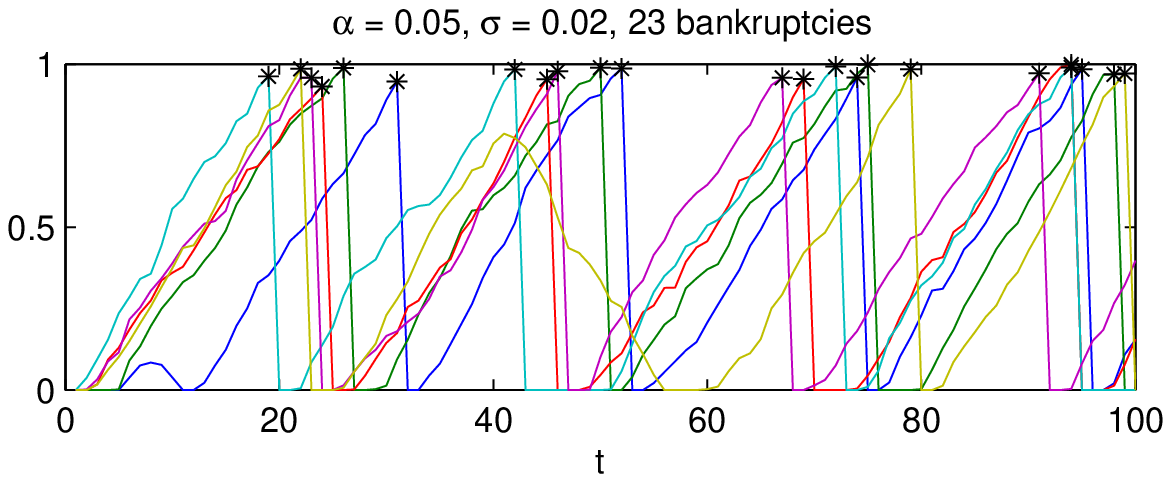}
  \end{center}
  \caption{Example of six trajectories of the stochastic process for
    $\phi$ with fixed trend strength $\alpha$ and decreasing noise level
    $\sigma$.  The number of failures first decreases but then increases.
    Since decreasing the noise level is equivalent to increasing the
    hedging level, the figure suggests that there is an optimal hedging
    level which minimizes the number of failures.}
  \label{fig:Runs}
\end{figure}

As an example, Figure \ref{fig:Runs} shows six trajectories of the
stochastic process defined in Eq. (\ref{eq:dynmf}) for a fixed value of
trend strength $\alpha$ and decreasing value of noise level $\sigma$.

In the following, we will investigate the role of noise on the
probability of failure by computing the pdf of $\phi$ in the limit of
large $t$, which represents the probability distribution of fragility in
the steady state of the process. Such pdf can be interpreted both as the
firm's individual probability of having a given value of fragility and as
an histogram of fragility values of an ensemble of firms.

\subsection{Dynamics of Fragility as Persistent Random Walk}
\label{subsec:PRW}
Since varying the hedging level $k$ is equivalent to varying the noise
level, in the following we definitely drop $k$ from Eq. (\ref{eq:dynmf})
and we study the process
\begin{equation}
  \phi(t+1) = \phi(t)+\sigma \xi(t) + \alpha\sign(\Delta\phi(t))
  \label{eq:dynmf}
\end{equation} 
Assuming that the boundary conditions are not effective during two
consecutive time steps, we can derive from \eqref{eq:dynmf} the
expression of $\phi(t+2)$ in terms of $\phi(t)$.
\begin{equation}
  \phi(t+2) = \phi(t)+\sigma(\xi(t+1)+\xi(t)) +
  \alpha\left[\sign(\Delta\phi(t))+
    \sign(\sigma\xi(t)+\alpha\sign(\Delta\phi(t)))\right].\label{eq:dyn2mf}
\end{equation}

Obviously, the last term in the square parentheses can only take the
values $-2,0$ or $2$, depending on the $\sign$ of $\Delta\phi(t)$ and the
probability
\[
\Prob(\sign(\sigma\xi(t)+\alpha\sign(\Delta\phi(t))) =
\sign(\Delta\phi(t))).
\]
This probability is
\[\Prob(\sigma\xi < \alpha) = \int_{-\infty}^{\alpha}f_{\sigma\xi}(x)dx
=\int_{-\infty}^{\frac{\alpha}{\sigma}}f_{\xi}(x)dx \] due to the
symmetry of $f_\xi$. We define $q(\alpha,\sigma) := \Prob(\sigma\xi <
\alpha)$ as the \emph{probability to keep the trend}. Denoting with
$F_\xi(x)$ the cumulative distribution function (cdf) of $\xi$ then it
holds $q(\alpha,\sigma) = F_\xi(\frac{\alpha}{\sigma})$.
We can then reformulate the process (\ref{eq:dynmf}) as
\begin{equation}
  \phi(t+1) = \phi(t) + \sigma\xi(t) + \alpha\tr(t)
  \label{eq:dyntr}
\end{equation}
where $\phi(t+1)$ is set to zero if it falls out of the interval
$[0,\theta]$. The function '$\tr$' is the discrete stochastic process
\begin{equation}
  \tr(t+1) = \eta\,\tr(t)  \qquad \textrm{ with } \eta = \left\{
    \begin{array}{cl}
      1  & \textrm{with probability $q$} \\
      - 1 & \textrm{with probability $1-q$}
    \end{array} \right.\label{eq:deftr}
\end{equation}
with possible initial values $\tr(0)=\{1,-1\}$ both with probability
$\frac{1}{2}$. Notice that $\tr$ is not affected when $\phi$ hits any of
the two thresholds. This implies that typically new firms are created
with positive trend. This hypothesis simplifies the analysis but does not
affect the result as discussed in Section \ref{sec:robustness}.

There are two important differences between the $\sign$-process
(\ref{eq:dynmf}) and the $\textrm{trend}$-process (\ref{eq:dyntr}).  The
first regards the behavior at the boundaries. Suppose both processes get
to 0 at time $t-1$ coming from a positive value at time $t-2$ and remain
at 0 at time $t$ (because, for instance, in the $\sign$-process
$\xi(t-1)$ and $\xi(t)$ were negative and in the $\tr$-process
$\eta(t-1)$ and $\eta(t)$ were $1$). In this case, the term
$\sign(\Delta\phi(t))$ in Eq.~(\ref{eq:dynmf}) is zero and therefore the
$\sign$-process will switch to a positive value at time $t+1$ with
probability $\frac{1}{2}$.  In contrast, the corresponding term $\tr(t)$
in Eq.~(\ref{eq:dyntr}) can never be zero (by definition its range is
$\{-1,+1\}$ and the $\tr$-process will switch to a positive value at time
$t+1$ with probability $1-q$. This means that when the noise $\sigma$ is
small and therefore $q$ is close to $1$, the $\tr$-process tends to stay
longer at 0, compared to the $\sign$-process.  The $\tr$-process can be
easily modified to better approximate the $\sign$-process by redefining
what happens at zero. We discuss possibile modifications and their
implications in Section \ref{sec:robustness}.

The second difference between the two processes concerns the dependencies
of the draws of the random variables. Eq.~(\ref{eq:dynmf}) implies that
$\sign(\Delta\phi(t))=\sign[\xi(t-1)+\alpha \sign(\Delta(\phi(t-1)))]$
and therefore $\xi(t-1)$ affects directly $\phi(t)$ and indirectly also
$\phi(t+1)$ through the term $\sign(\Delta\phi(t))$. In contrast, in the
$\tr$-process the term $\tr(t)$ evolves independently of the draws of the
random variable $\xi$

We now compare the $\tr$-process with a process called persistent random
walk (PRW) in the physics literature. PRW is a variant of the classic
random walk in which the walker has a probability $q$ to keep the
direction and $1-q$ to switch direction. If we neglect the noise term
$\sigma\xi(t)$ in (\ref{eq:dyntr}) and start with $\phi(0)=0$, then
$\phi$ evolves like a PRW on $\mathbb{Z}$. The PRW obeys the
telegrapher's equation in the continuous limit
\cite{Araujo.ea91,Masoliver91,Weiss02}. An important property of the PRW
is that it deviates, in a transient phase, from the linear scaling of the
variance of the displacement with time, $<x^2> \sim t$ that is
characteristic of the RW. Indeed, starting with all probability mass in
zero, the variance first increases quadratically, $<x^2> \sim t^2$, due
to waves that start towards $-\infty$ and $+\infty$ (ballistic scaling).
After a continuous transition, the variance grows linearly as in the
usual RW (diffusive scaling) and in the limit of large $t$, it evolves as
$\frac{q}{1-q}t$. Therefore, if $q$ is close to $1$, the variance grows
still linearly for large $t$, although with a high diffusion coefficient
$\frac{q}{1-q}$.  Compared to a pure persistent random walk, our process
includes, additionally, a continuous additive noise, a sort of reflecting
lower bound at zero, an absorbing bound $\theta$ (which leads to a
rebirth of firms with zero fragility), and the fact that the probability
$q$ of keeping the trend depends monotonously on $\frac{\alpha}{\sigma}$.

\subsection{Dynamics on the probability density function of $\phi$}
\label{subsec:pdf_dynamics}

In order to estimate the probability that the fragility $\phi$ hits the
treshold $\theta$, we want to know how its pdf evolves in time, and in
particular to estimate its stable pdf if this exists.

However, it is important to notice that, at any time step $t$, the state
of the process (\ref{eq:dyntr}) is determined both by the value of
$\phi(t)$ and by the value of the trend $\tr(t)$ which evolves as the
simple two-state process (\ref{eq:deftr}).

In order to study the evolution of the pdf of $\phi$ one has to study the
evolution of the pdf of the whole process (\ref{eq:dyntr}-\ref{eq:deftr})


Since the trend process takes only two values, we can divide the pdf of
$\phi(t)$ into two parts, corresponding to negative trend ($\tr(t)=-1$)
and positive trend ($\tr(t)=+1$). We define the two functions as
$f_{\phi(t)}^{-}: [0,\theta] \rightarrow \R_{\geq 0}$ and
$f_{\phi(t)}^{+}: [0,\theta] \rightarrow \R_{\geq 0}$.

The pdf of the whole process is determined by the pair of functions
$(f_{\phi(t)}^-,f_{\phi(t)}^+)$ under the condition that $\int_0^{\theta}
f_{\phi(t)}^{-}(\phi') + f_{\phi(t)}^{+}(\phi') d\phi' = 1$.

From this pair of functions we can derive the pdf of $phi$ as
$f_{\phi(t)}= f_{\phi(t)}^{-} + f_{\phi(t)}^{+}$. In other words,
$\int_{\phi'}^{\phi'+d\phi'} f_{\phi(t)}^{-}(\phi') d\phi'$ represents
the probability to have fragility in $[\phi, \phi+d\phi]$ and at the same
time a downward trend, $\tr(t)=-1$. Analogous relation holds for the
positive trend.

It is also possible to derive the pdf of $\tr$ as $f_{\tr(t)}=(\int
f_{\phi(t)}^{-}, \int f_{\phi(t)}^{+})$, which is a pair of scalar values
specifying the probability of having negative and positive trend and
which is therefore not really a pdf but a probability mass function
defined on $\{ -1,+1\}$.

We also define $\delta_\alpha$ to be the Dirac $\delta$-distribution with
mass shifted by $\alpha$ (also known as $\delta(\cdot-\alpha)$), '$\ast$'
to be the convolution operator for functions (defined for two functions
$h_1,h_2:\R\to\R$ as $(h_1\ast h_2)(\varphi) = \int
h_1(y)h_2(\varphi-y)dx$), $f_{\sigma\xi}$ to be the pdf of the noise).

\begin{proposition} 
  Let the pdf of $\phi(t)$ be $(f_{\phi(t)}^-,f_{\phi(t)}^+)$. If the
  stochastic evolution of $\phi$ evolves as defined in Eq.
  \eqref{eq:dyntr}, then the pdf of $\phi(t+1)$ is
  $(f_{\phi(t+1)}^-,f_{\phi(t+1)}^+)$ with
  \begin{align}\nonumber
    f_{\phi(t+1)}^{-} & = g_{t}^{-}\one_{[0,\theta]} + (b^-(t) +
    z^-(t))\delta_0 \\ \label{eq:dynpdf2} f_{\phi(t+1)}^{+} & =
    g_{t}^{+}\one_{[0,\theta]} + (b^+(t) + z^+(t))\delta_0.
  \end{align}
  The functions $g_t^-,g_t^+$ are defined as
  \begin{align}\nonumber
    g_{t}^{-} = (qf_{\phi(t)}^{-}\ast\delta_{-\alpha}
    +(1-q)f_{\phi(t)}^{+}\ast\delta_\alpha)\ast f_{\sigma\xi} \\
    \label{eq:dynpdf1} g_{t}^{+} =
    ((1-q)f_{\phi(t)}^{-}\ast\delta_{-\alpha}
    +qf_{\phi(t)}^{+}\ast\delta_\alpha)\ast f_{\sigma\xi}
  \end{align}
  and $b^-(t) = \int_{\theta}^{+\infty}g_{t}^{-},\ b^+(t) =
  \int_{\theta}^{+\infty}g_{t}^{+}$ are the \emph{probabilities to go
    above $\theta$} for $\phi(t)$ with negative or, respectively,
  positive trend, and $z^-(t) = \int_{-\infty}^{0}g_{t}^{-},\quad z^+(t)
  = \int_{-\infty}^{0}g_{t}^{+}$ are the \emph{probabilities to go below
    zero} for $\phi(t)$ with negative or, respectively, positive trend.
  \label{prop:dynam-prob-dens}
\end{proposition}

\begin{figure}[htb]
  \centering
  \includegraphics[width=\textwidth]{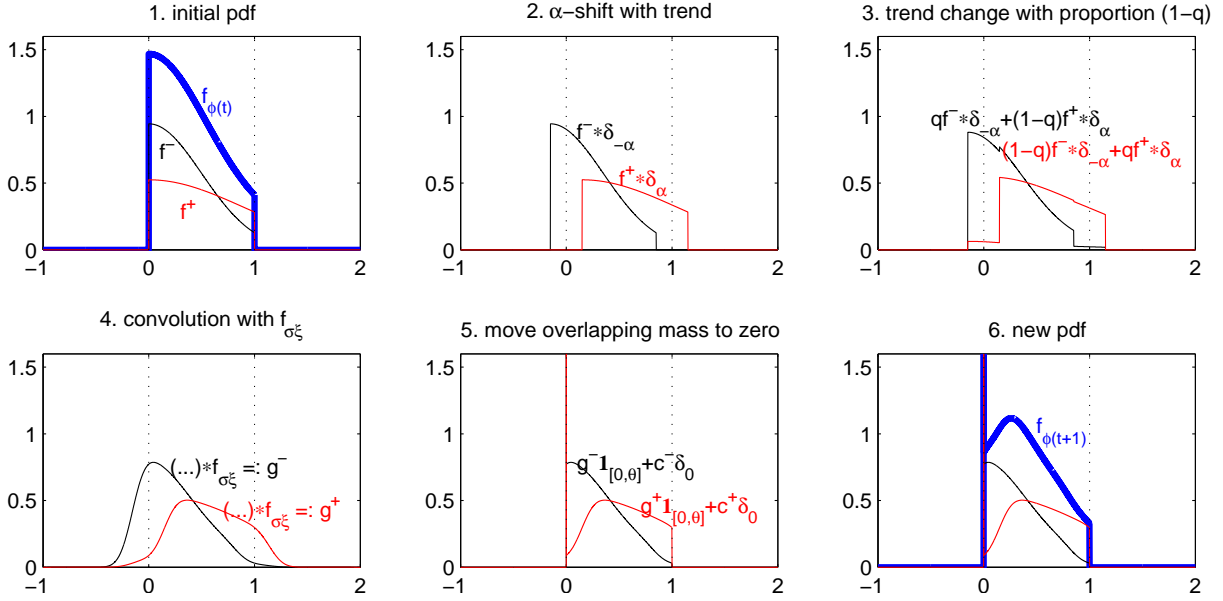}
  \caption{Shifting, trend switching, shock addition and redistribution
    of overlapping mass for given $f_{\phi(t)} = f^-_{\phi(t)} +
    f^+_{\phi(t)}$. Parameters used: $\theta = 1$, $\alpha = \sigma =
    0.15$, the noise pdf $f_{\sigma\xi}$ is gaussian. This choice implies
    $q \approx 0.8413$.  In step 5 $c^- = (b^-(t)+z^-(t))\ ,c^+ =
    (b^+(t)+z^+(t))$.}
  \label{figExplain}
\end{figure}

The successive steps in the computation of $g_t^-$ and $g_t^+$ are
illustrated in Figure \ref{figExplain} (steps 2 to 4), while the
computation of $(f_{\phi(t+1)}^{-},f_{\phi(t+1)}^{+})$ is illustrated in
step 5.

\begin{proof}
  First, we look at (\ref{eq:dynpdf2}). Notice that the convolution
  operation is commutative and distributive with respect to the operation
  of addition. Thus, the order of the computation does not matter.

  Adding the noise term $+\sigma\xi(t)$ to $\phi(t)$ in
  Eq.~(\ref{eq:dyntr}), corresponds to the convolution of the pdf of
  $\phi(t)$ with the pdf of the noise $f_{\sigma\xi}$.

  The term $\alpha\tr(t)$ in the same equation, implies that the part of
  the pdf representing the upward trend is shifted upwards by $\alpha$
  and that the part representing the downward trend is shifted downwards
  by $\alpha$. This is because shifting a function along the $x$-axis is
  represented by convolution with a shifted delta-function.

  If the process is on a downward trend, it will keep that trend with
  probability $q$ and switch with probability $(1-q)\,$. The vice-versa
  holds for the upward trend. Thus, a $q$-fraction of $f_{\phi(t)}^{-}$
  will remain in $f_{\phi(t+1)}^-$, while a $(1-q)$-fraction of
  $f_{\phi(t)}^{+}$ will join $f_{\phi(t+1)}^-$. The vice-versa holds for
  $f^+$.

  Finally, Eq. (\ref{eq:dynpdf1}) ensures that all probability mass which
  overlaps the interval $[0,\theta]$ is distributed back to $[0,\theta]$.
  The overlapping probability mass is determined by
  $b^-(t),b^+(t),z^-(t),z^+(t)$ and according to the boundary conditions,
  it is put in a $\delta$-peek at zero, while the trend information gets
  conserved.
\end{proof}

Notice that other definitions for rebirth after failure can easily be
modeled by changing $\delta_0$ in Eq. (\ref{eq:dynpdf2}) to any other pdf
(for example to the pdf of the uniform distribution if firms should be
reborn with random and equally distributed fragility). Further on, also
other rules for changes of the trend can be modeled by replacing $(b^-(t)
+ z^-(t))$ and $(b^+(t) + z^+(t))$ by other combinations.

To better approximate the $\sign$-process, one should replace $z^-(t)$
and $z^+(t)$ by $\frac{1}{2}(z^-(t) + z^+(t))$. This models the fact that
a firm with fragility zero for two time steps has a zero trend, and
switches with equal probability to the upward or downward trend,
regardless of the former trend.

Given an initial pdf $(f^-_{\phi(0)},f^+_{\phi(0)})$,
Proposition \ref{prop:dynam-prob-dens} defines a time-discrete evolution
of the probability density function of the firm's fragility.

In the following of this section, we will use the dynamics as defined in
\eqref{eq:dynpdf2}.

\begin{proposition}
  Consider the process defined in Eq. (\ref{eq:dyntr}), where $\xi$ is a
  normally distributed random variable with mean zero, variance one and
  pdf $f_\xi$, with noise level $\sigma>0$, trend strength $\alpha \geq
  0$, failing threshold $\theta$.

  If $q(\alpha,\sigma) = \Prob(\sigma\xi<\alpha) < 1$, then there exists
  a unique stable pdf $(f^-_\ast,f^+_\ast)$.
  
  Furthermore, any initial pdf $(f^-_{\phi(0)},f^+_{\phi(0)})$ converges,
  under the evolution defined in Proposition \ref{prop:dynam-prob-dens},
  to $(f^-_\ast,f^+_\ast)$ geometrically fast, with $\int f^-_\ast = \int
  f^+_\ast = \frac{1}{2}$.
\end{proposition}

\begin{proof}
  We want to apply a theorem known as Birkhoff-Jentzsch Theorem
  \cite[Page 224, Theorem 3]{birkhoff1957ejs}. It is an extension of the
  famous Perron-Frobenius Theorem for nonnegative matrices to
  infinite-dimensional vector spaces.
  
  It is easy to see that, for any bounded pdf
  $(f^-_{\phi(t)},f^+_{\phi(t)})$ the two parts of the pdf
  $(f^-_{\phi(t+1)},f^+_{\phi(t+1)})$ are continuous on $]0,\theta]$,
  have a $\delta$-peak at zero and full support $[0,\theta]$. So, after
  one iteration the dynamics (\ref{eq:dynpdf2}) remain in the space of
  pairs of bounded continuous functions with a $\delta$-peaks at zero.

  Let us define the operator $P$ on the vector space of these functions
  such that it transforms $(f^-_{\phi(t)},f^+_{\phi(t)})$ into
  $(f^-_{\phi(t+1)},f^+_{\phi(t+1)})$. This operator fulfills the
  conditions of the Birkoff-Jentsch Theorem: it is in fact a uniformly
  positively bounded linear operator.

  It is bounded because, trivially, the integral of the pdf is always
  one. The linearity is also easily checked since all entities in the
  definition of the dynamics Eqs. (\ref{eq:dynpdf2}-\ref{eq:dynpdf1}) are
  linear.

  Now we show that it is also uniformly positive (as defined in
  \cite[Page 219]{birkhoff1957ejs}). In our case an eigenvalue of $P$
  must be $\lambda = 1$. As lower bound for
  $(f^-_{\phi(t+1)},f^+_{\phi(t+1)})$ we take $(e,e)$ with $e =
  (1-q)c_1(\one_{[0,\theta]} + \delta_0)$ with $c_1 =
  f_{\sigma\xi}(\theta+\alpha)$. This is obviously the lowest value
  $(f^-_{\phi(t+1)},f^+_{\phi(t+1)})$ can take after one iteration
  because of convolution with $f_{\sigma\xi}$. (Take e.g.
  $(f^-_{\phi(t)},f^+_{\phi(t)}) = (0,\delta_\theta)$ as a 'worst case'.)
  Further on, an upper bound exists $c_2(\one_{[0,\theta]} + \delta_0)$
  with $c_2 = f_{\sigma\xi}(0)$. Thus, there exists the desired strech
  parameter $K=c_2$ for the Birkoff-Jentsch Theorem.

  The Birkoff-Jentsch Theorem now states that there is a unique
  $(f^-_\ast,f^+_\ast)$ and that for any inital pdf convergence to
  $(f^-_\ast,f^+_\ast)$ happens by iteration of the operator $P$
  geometrically fast.

  The equations $\int f^-_\ast = \int f^+_\ast = \frac{1}{2}$ are
  obvious, because any other distribution of mass in the parts of the pdf
  would not stay constant because of the equal exchange of $(1-q)$
  fractions in each step.
\end{proof}

This is probably not the most general form of the theorem. Other forms of
$f_{\sigma\xi}$ than normal (even with bounded support) also often lead
to stabilization. But a proof is not that straight forward.

If we exchange the terms $z^-(t)$ and $z^+(t)$ by $\frac{1}{2}(z^-(t) +
z^+(t))$ to better approximate the $\sign$-process, then fractions of
mass in the parts of the still existing unique stable pdf will not be
equal anymore.

From this section we conclude that there is a unique attractive stable
distribution for the probability density of fragility in the $\tr$
process of Eq.~\ref{eq:dyntr}. Moreover, the probability to fail at time
$t$
\begin{equation}
  \label{eq:failprob}
  b(t) = b^-(t) + b^+(t)
\end{equation}
converges to fixed value which we define as the \emph{limit failure
  probability}.
\begin{equation}
  \label{eq:sysrisk}
  b^\ast = \lim_{t\to\infty} b(t).
\end{equation}
 
\section{Numerical results}
\label{sec:numerical_results}

Unfortunately, the unique stable pdf $(f^+_\ast,f^-_\ast)$ seems not to
have a closed form, or at least not an easy one. Therefore, we compute it
numerically. We set $\theta=1$ (without loss of generality) and $f_{\xi}$
to be Gaussian (with mean zero and variance one) and we explore the
$(\alpha,\sigma)$-parameter space. Each pair of values $(\alpha,\sigma)$
corresponds to a value of $q$ which lies in the interval $[0.5,1]$.
Notice that, assuming a different pdf for the noise would imply different
values of $q$ (cf. Section \ref{sec:robustness}).

\begin{figure}[htb]
  \begin{center}
    \mbox{$\alpha = 0.1, \sigma = 0.3, q \approx 0.6306$} \\
    \includegraphics[width=\textwidth]{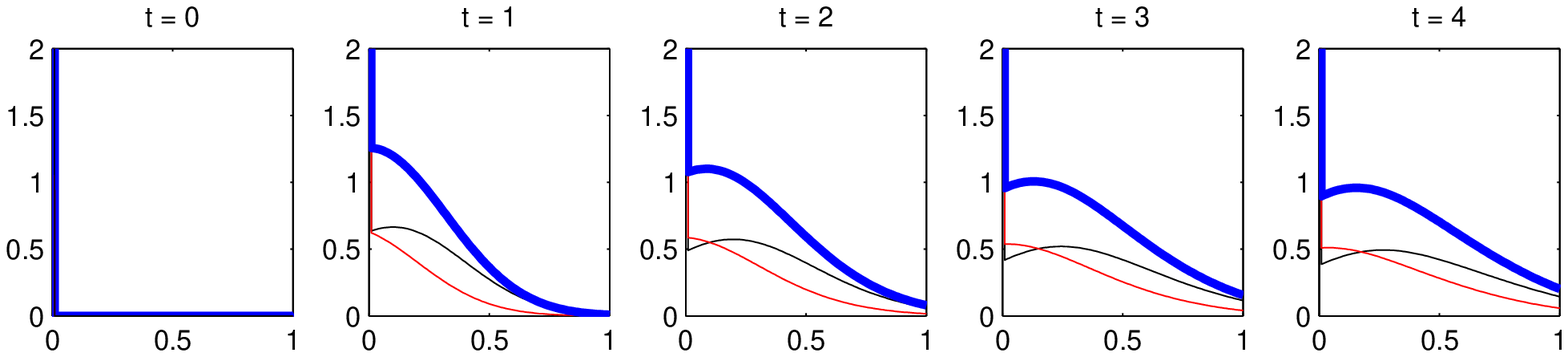}
    \vspace{0.4cm}
    \mbox{$\alpha = 0.2, \sigma = 0.15, q \approx 0.9088$} \\
    \includegraphics[width=\textwidth]{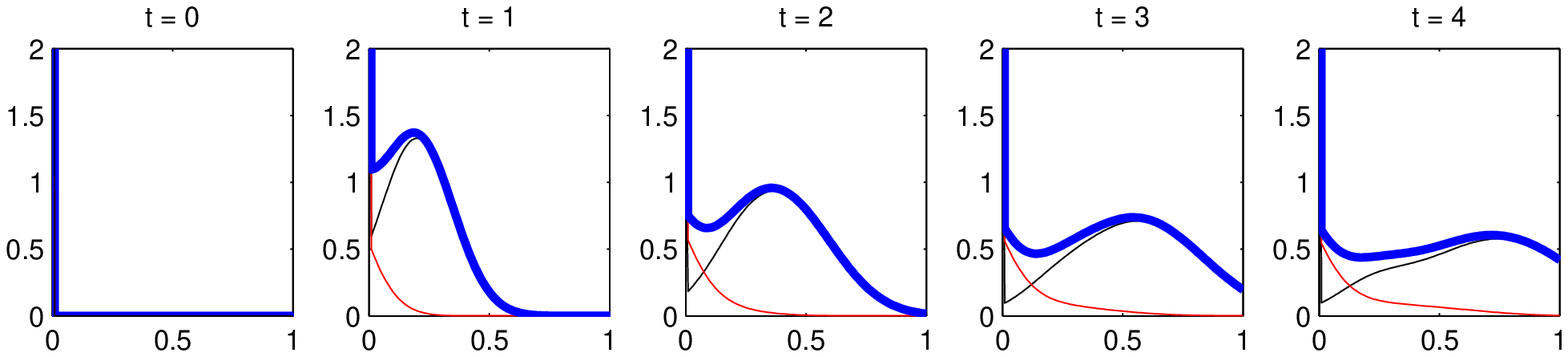}
    \vspace{0.4cm}
    \mbox{$\alpha = 0.3, \sigma = 0.1, q \approx 0.9987$} \\
    \includegraphics[width=\textwidth]{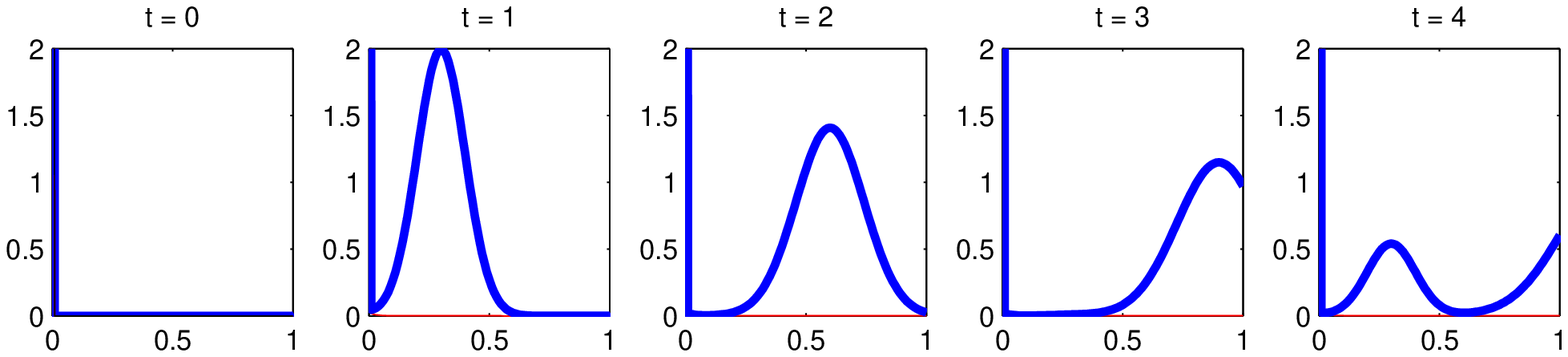}
    \caption{The first four time steps with initial pdf $f_{\phi(0)} =
      f^+_{\phi(0)} + f^-_{\phi(0)} = \frac{1}{2}\delta_0 +
      \frac{1}{2}\delta_0$ and different parameters, $f_\xi$ is Gaussian,
      the $q$-values are computed from $\alpha$ and
      $\sigma$.}\label{fig:process}
  \end{center}
\end{figure}

\begin{figure}[htb]
  \centering
  \includegraphics[width=\textwidth]{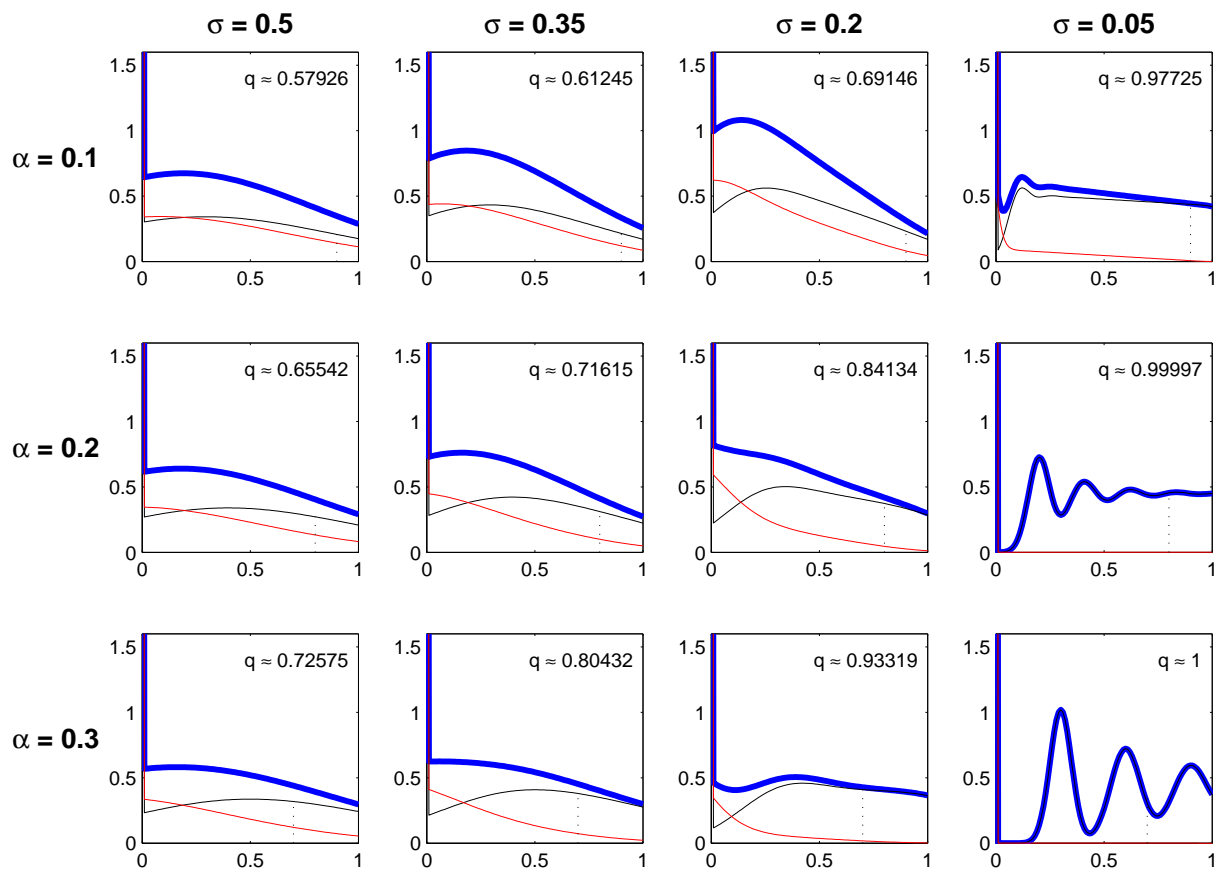}
  \caption{Stable pdfs for selected trend strength $\alpha$ and noise
    level $\sigma$ values.}
  \label{fig:stable1}
\end{figure}

Figure \ref{fig:process} shows the first time steps of the pdf evolution
for different $(\alpha,\sigma)$ values. Here the initial value of
fragility is zero and the initial value of the trend is $\pm1$ with equal
probability. Therefore, the initial pdf is $(f^+_{\phi(0)},f^-_{\phi(0)})
= \frac{1}{2}(\delta_0,\delta_0)$.
The parameter choice in the first row of plots in Figure
\ref{fig:process} corresponds to a relatively low trend strength $\alpha$
compared to the noise level $\sigma$ and thus to a value of $q$ only
slightly above its minimum $0.5$. The random term $\sigma\xi$ plays the
major role in the process and in this regime the persistent random walk
behaves similar to the usual random walk. This leads to a fast
convergence of the pdf: after only four time steps (last plot in the
row), the pdf is already close the stable pdf (cf. Figure
\ref{fig:stable1}). Notice that there is a significant delta peak at 0
(going beyond limit of the ordinate axis in the plot) which collects the
probability to go below 0 and the probability to go above 1.

In the second row of plots in Figure \ref{fig:process}, the values of
$(\alpha,\sigma)$ correspond to values of $q$ closer to one. This implies
that most of the mass of the probability density function corresponding
to the downward trend ($f^-$) stays close to zero. On the other hand, the
mass in $f^+$ moves with a wave towards the failure threshold (which is
at 1, since the abscissa represents $\phi$ and $\theta=1$). The wave
smoothes out due to the repeated convolution with $f_{\sigma\xi}$.
Finally, in the third row of plots in Figure \ref{fig:process} $q$ is
very close to one. In this case the wave towards the failure threshold
repeats several times until it smoothes out. Notice that in the limit
$\sigma\to 0$, and thus $q\to 1$ (not shown in the figure), the pdf of
$\phi$ will not converge. There will be a delta peak which moves
constantly upwards (modulo the redistribution of its mass in zero).

Figure \ref{fig:stable1} shows instead the stable pdfs for some specific
values of $\alpha$ and $\sigma$. The pdf's were computed by iteration of
Eq. (\ref{eq:dynpdf2}) with initial uniform distribution on $[0,\theta]$
and discretization of the interval $[0,1]$ in steps of $0.01$. We
proceeded until the norm of the difference in one time step was smaller
than an accuracy level of $10^{-6}$. There were no hints that a finer
discretization would improve the result.

The figure shows that the stable pdf is approximatively linearly
decreasing for high values of fragility (except for the wavy pdf's
obtained with high $\alpha$ and low $\sigma$). The slope of the linear
decrease is non-monotonously controlled by $\sigma$ and $\alpha$. It is
easy to explain the slope in some cases, although this is not the case in
general. When $q$ is close to 1, it is very unlikely that a trajectory of
the process switches direction. A trajectories with positive trend moves
steadily along the whole range of values $[0,1]$, repetitively hits the
threshold 1 and gets reset to 0. In contrast a trajectory with negative
trend reaches 0 and stays there. As a result, $f^+$ tends to a uniform
distribution in $[0,1]$ and $f^-$ tends to a delta peak in 0. On the
other hand, $q$ close to $0.5$ is implied by $\sigma$ much larger than
$\alpha$. In this regime, $\phi$ diffuses very fast which leads again to
a rather flat distribution for both $f^-$ and $f^+$. In contrast, for
intermediate values of $q$ (for instance $\alpha=0.1, \sigma=0.2$), the
profile has a pronounced negative slope for high $\phi$.

In the regime of high $\alpha$ and $\sigma$ close to 0, the trajectory
evolves by almost discrete jumps of magnitude close to $\alpha$. This
results in a wavy stable pdf with peaks at multiples of $\alpha$. But the
wavy pattern oscillates around a line with flat slope, which is consistent
with what found in the case of high $q$ and $sigma$ not too close to 0.

We are most interested in the limit failure probability which is our
proxy for the systemic risk. It depends on trend strength $\alpha$ and
noise level $\sigma$. So, we computed $b^\ast = b^\ast(\alpha,\sigma)$
for the parameter set $\alpha,\sigma \in ]0,0.5]$.

Figure \ref{fig:systemic} shows that for fixed trend strength $\alpha$
there is an intermediate optimal $\sigma$ which leads to minimal systemic
risk. In contrast, for a fixed noise level $\sigma$ there is no
intermediate minimum when varying the trend strength $\alpha$. Raising
the trend strength always increases the systemic risk. The lines for high
$\sigma$ and low $\sigma$ intersect. This resembles the existence of the
intermediate optimum for fixed $\alpha$.
\begin{figure}[htb]
  \centering
  \includegraphics[width=\textwidth]{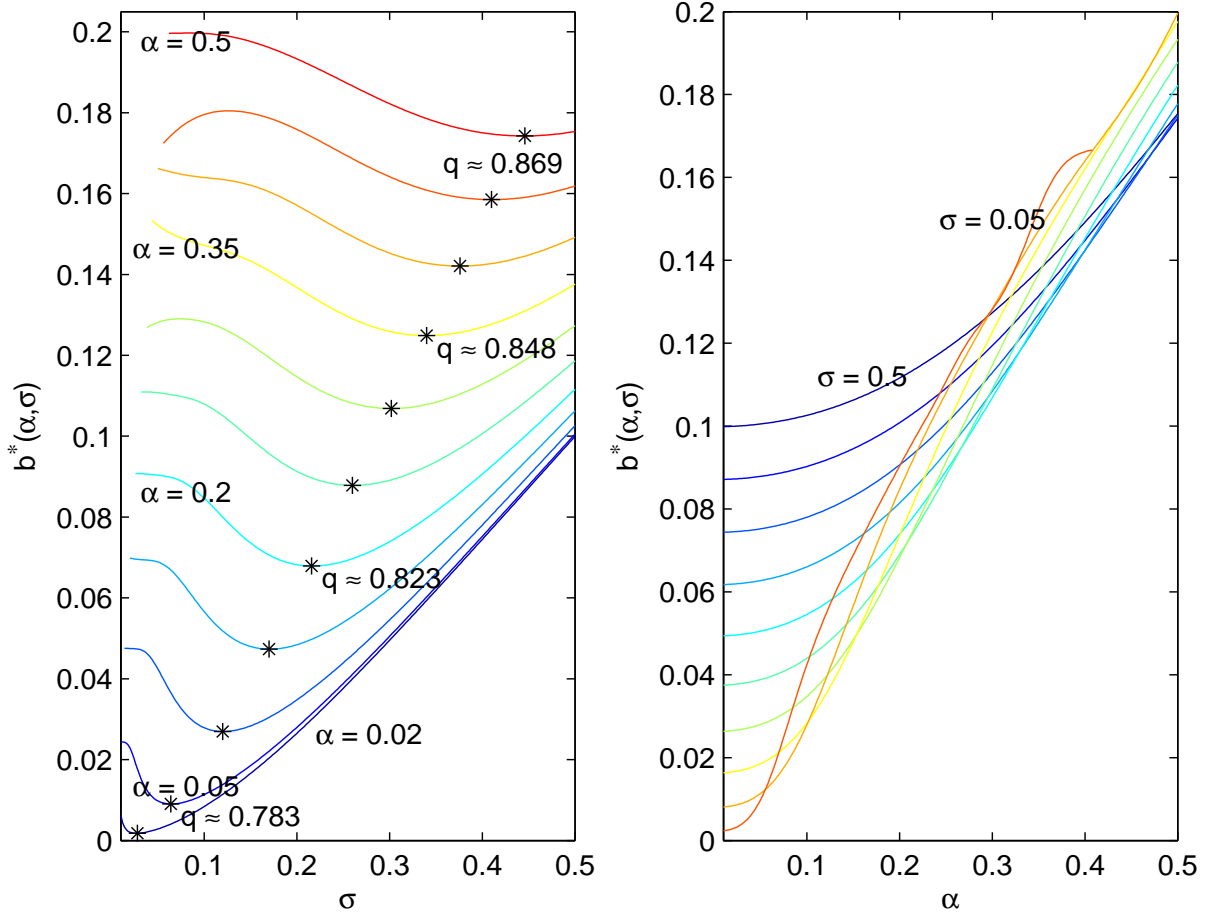}
  \caption{The limit failure probability with respect to trend strength
    $\alpha$ and standard deviation of shocks $\sigma$. (Noncomplete
    lines are due to extremely long convergence times.)}
  \label{fig:systemic}
\end{figure}
The left plot in Figure \ref{fig:systemic} shows also values of the
probability to keep the trend $q$ at the intermediate minima of the limit
failure probability with respect to $\sigma$, given a fixed trend
strength $\alpha$. It turns out that the optimal noise level lies at a
value of $q$ roughly between $0.75$ and $0.9$. The value of $q$
corresponding to the local minimum decreases slowly with $\alpha$. This
is better visible in Figure \ref{fig:systemic} where we take a bird eye's
view on the $(\alpha,\sigma)$-plane, where the level lines of equal $q$
appear as rays from the origin. The ordinate represents $q=0.5$, the
abscissa $q=1$.
 
\begin{figure}[htb]
  \centering
  \includegraphics[width=0.7\textwidth]{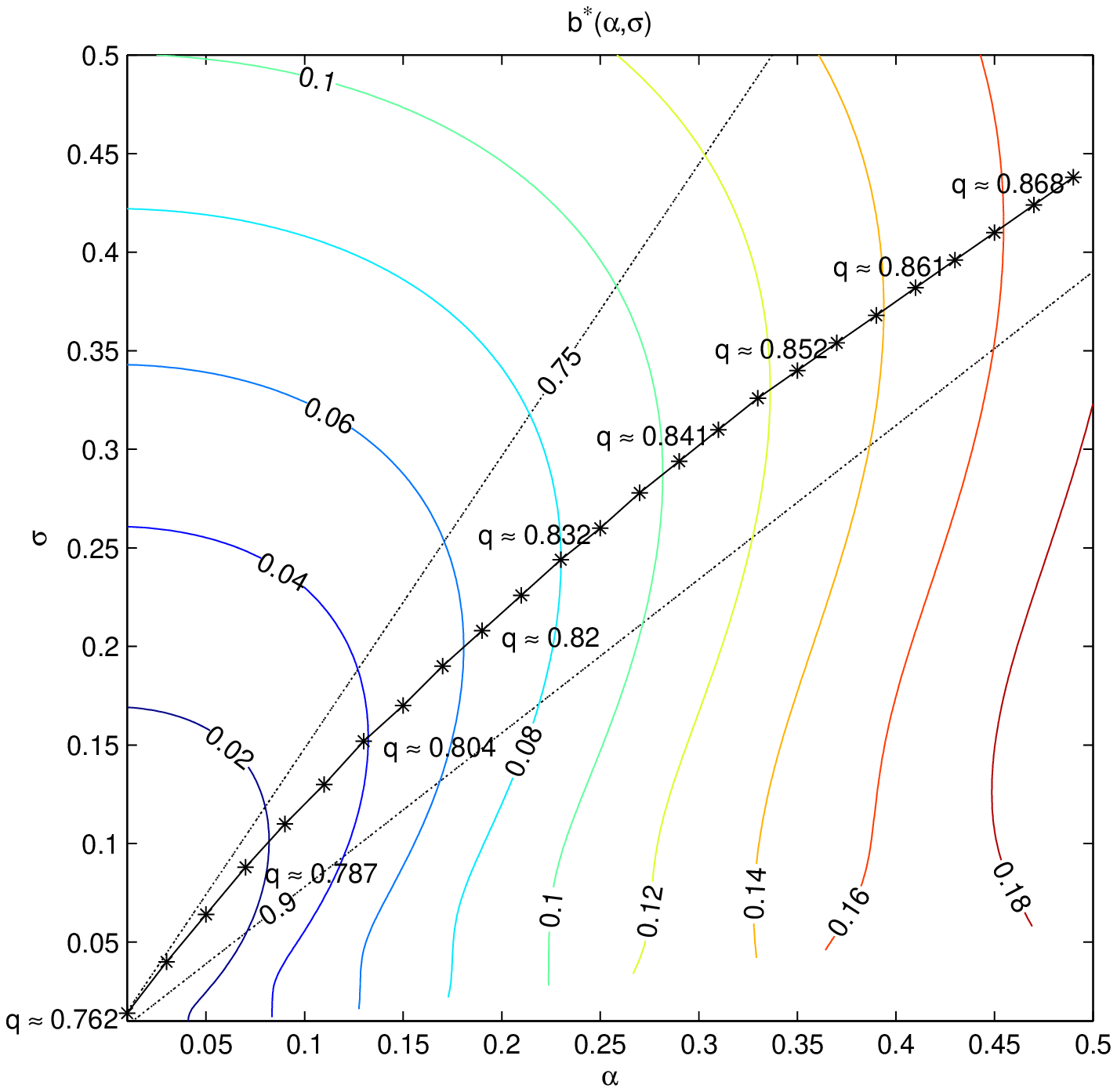}
  \caption{Contour plot for the limit failure probability
    $s(\alpha,\sigma)$ and contour lines for $q(\alpha,\sigma)=0.8,0.9$.
    The solid black line denotes the optimal value of $\sigma$ regarding
    a fixed $\alpha$. }
  \label{fig:systemic}
\end{figure}

\section{Robustness of results}
\label{sec:robustness}

We checked other pdf's for the noise besides the Gaussian and in most
cases we also observe convergence to a unique stable pdf. Notice that
convergence is not assured in general by Proposition
\ref{prop:dynam-prob-dens}. We observed quantitative changes in the
results but not qualitative ones in the sense that there always exists an
optimal noise level for a fixed trend strength.

In our model, firms fail when their fragility hits a threshold and are
recreated with an initial value of fragility zero and an initial trend
proportional to the number of failing firms with that trend (so mostly
with upward trend). This is a strong assumption and therefore we checked
three other scenarios, in particular to test whether the phenomena of an
intermediate optimal noise level is robust against these modifications.
\begin{itemize}
\item If a new born firm is assigned a positive or negative trend with
  equal probability (instead of proportional to the number of failing
  firms of that trend) then the probability to have a positive trend
  $\int f^+_{\phi(t)}$ converges to a fixed number below $\frac{1}{2}$
  which depends on $q$. In the extreme case, $q=1$ it goes to zero. That
  would implies that the probability to fail will also go to zero in the
  limit. We saw that for a fixed trend strength there is a critical noise
  level that implies such high $q$ that the systemic risk drops to zero
  when the noise level gets below. Nevertheless, for low trend strength
  values ($\alpha$ below about 0.12) an intermediate optimal noise level
  still exists until further decreasing the noise level causes the sudden
  drop due to the extinction of the upward trend.  One may criticize this
  variation of the model because it does not converge to equal
  proportions of positive and negative trend. But stable equal
  probabilities for upward and downward trend seems quite reasonable
  because judgement of fitness is always done comparatively in an
  economy. If economy divides firms in good and bad ones this should not
  lead to a possible die out of one class.

\item Another suggestion against our original model could be that firms
  are not born with zero fragility but i.e. random and uniformly
  distributed in the fragility interval. This obviously changes the limit
  pdf, but at least in this example the qualitative behavior with the
  existence of an optimal number of hedging partners for given rend
  strength remains the same.

\item Another idea is to renormalize the probability mass after a
  failure. We do this as follows: we do not redistribute the probability
  mass after a failure to zero but just rescale $f^+$ proportional to its
  actual shape such that it has the same total amount as before. The same
  with $f^-$. On the level of individual firms this means that new firms
  are born with fragilities drawn randomly from the actual distribution
  of fragilities with that trend. That means if the distribution of
  fragilities is double peaked new firms are most likely to appear with
  fragilities around that two peaks. This dynamics imply that a given
  peak structure gets amplified by the evolution of new firms. In fact
  this dynamic fragility distribution for new firms leads to an
  amplification of mass in high fragility intervals. That means that with
  high probability new firms are born with high fragility (which seems
  reasoable). In the limit these regimes are characterized by virtually
  all firms with positve trend failing each year. That means that
  decreasing the noise level (which increases $q$) is even more
  dangerous. Nevertheless, there still exists an intermediate optimal
  noise level for a given trend strength to minimize the systemic risk.
\end{itemize}

\section{Conclusions}
\label{sec:conclusions}

We have presented a simple model for the stochastic evolution of the
fragility of units in a network. The model applies in particular to
networks of firms connected via financial relationships. The basic
ingredients of the model consist in a mechanism of risk sharing that
leads to decrease the fluctuation of the fragility and in a mechanism of
reinforcing feedback on the fragility from the trend in the immediate
past of the fragility of the firm itself and its neighbors. Under this
assumptions, the number of bankruptcies in the system is minimized for an
intermediate density of links in the network e.g. for an intermediate
number of hedging partners. The result is of interest from the point of
view of policy design for the control of systemic risk.

The effect depends strongly on a dynamics divisions of firms into two
classes: the good evolving (with decreasing fragility) and the bad
evolving firms (with increasing fragility).  One might question that this hard cut
between the two classes exists. But we argue that actually, slight
differences in performance are exacty what investors like hedge funds
search for when they try to profit from
investments indepently of the economic trend. So, even very slight
differences may matter a lot for reinforcing trends. Further on, these
kind of investment strategies have become more popular. 

With respect to the original model, the analysis presented here neglects
the process of cascades of failures and therefore underestimates the
number of joint failures. However, its advantage is that the evolution of
the probability distribution of failures can be expressed analitically
and that the stable distribution (which we prove to exist and be unique)
can be computed numerically.

The impact of heterogeneity in the topology of the network is not studied
at this stage. Furthermore, the hedging network is not dynamic. This
implies for instance that firms do not have the possibility to interrupt
hedging relations with partner who do not perform well. This assumption
is certainly not very realistic on a time scale of years. However, it is
also true that many partnership or insurance contracts cannot be modified
in a very short time. Furthermore, in future work the impact of heterogeneous trend
strength, noise level and failing threshold should be studied.

\subsection*{Acknowledgements}
This work is part of a project within the COST Action P10 ``Physics of
Risk''. We appreciate financial support from the Swiss National Science
Foundation under the contract number C05.0148.  We thank Ulrich Krause
(Bremen) for pointing us to the Birkhoff-Jentzsch Theorem. We also thank
Mauro Napoletano for suggestions and helpful discussions.

\bibliographystyle{plain}

\bibliography{refs}

\end{document}